\documentclass[twocolumn]{aastex61}

\received{June 9, 2017}
\revised{June 14, 2017}
\accepted{June 14, 2017}
\submitjournal{ApJ}

\shorttitle{A lensed ultrabright submillimeter galaxy at $z=2.0439$}
\shortauthors{A. D\'{\i}az--S\'anchez et al.}
\usepackage{xcolor}
\begin{document}

\title{Discovery of a lensed ultrabright submillimeter galaxy at $z=2.0439$}

\correspondingauthor{A. D\'{\i}az--S\'anchez
  }
\email{Anastasio.Diaz@upct.es}

\author{A. D\'{\i}az--S\'anchez
}

\affil{
Departamento F\'\i sica Aplicada, Universidad Polit\'ecnica de Cartagena, Campus Muralla del Mar, 30202 Cartagena, Murcia, Spain\\
}

\author{S. Iglesias-Groth}
\affiliation{
Instituto de Astrof\'\i sica de Canarias,  V\'\i a L\'actea, La Laguna 38200, Spain\\
}
\affiliation{
Departamento de Astrof\'\i sica de la Universidad de La Laguna, Avda. 
Francisco S\'anchez, La Laguna, 38200, Spain
}

\author{R. Rebolo}
\affiliation{
Instituto de Astrof\'\i sica de Canarias,  V\'\i a L\'actea, La Laguna 38200, Spain\\
}
\affiliation{
Departamento de Astrof\'\i sica de la Universidad de La Laguna, Avda. 
Francisco S\'anchez, La Laguna, 38200, Spain
}
\affiliation{
Consejo Superior de Investigaciones Cient\'\i cas, Spain \\
}
\author{H. Dannerbauer}
\affiliation{
Instituto de Astrof\'\i sica de Canarias,  V\'\i a L\'actea, La Laguna 38200, Spain\\
}
\affiliation{
Departamento de Astrof\'\i sica de la Universidad de La Laguna, Avda. 
Francisco S\'anchez, La Laguna, 38200, Spain
}

%% Note that the \and command from previous versions of AASTeX is now
%% depreciated in this version as it is no longer necessary. AASTeX 
%% automatically takes care of all commas and "and"s between authors names.

%% AASTeX 6.1 has the new \collaboration and \nocollaboration commands to
%% provide the collaboration status of a group of authors. These commands 
%% can be used either before or after the list of corresponding authors. The
%% argument for \collaboration is the collaboration identifier. Authors are
%% encouraged to surround collaboration identifiers with ()s. The 
%% \nocollaboration command takes no argument and exists to indicate that
%% the nearby authors are not part of surrounding collaborations.

%% Mark off the abstract in the ``abstract'' environment. 
\begin{abstract}

We report an ultra-bright lensed submillimeter galaxy (SMG) at $z=2.0439$, {\it WISE} J132934.18+224327.3, identified as a result of a full-sky cross-correlation of the {\it AllWISE} and {\it Planck} compact source  catalogs aimed to search for bright analogs of the submillimeter galaxy SMMJ2135, the Cosmic Eyelash. Inspection of archival SCUBA-2 observations of the candidates revealed a source with fluxes (S$_{850 \mu m}$= 130 mJy) consistent with the {\it Planck} measurements. The centroid of the SCUBA-2 source coincides within 1 arcsec with the position of the {\it AllWISE} mid-IR source, and, remarkably, with an arc shaped lensed galaxy in {\it HST} images at visible wavelengths. Low-resolution rest-frame UV-optical spectroscopy of this lensed galaxy obtained with 10.4 m GTC reveals the typical absorption lines of a starburst galaxy. Gemini-N near-IR spectroscopy provided a clear detection of H$_{\alpha}$ emission. The lensed source appears to be gravitationally magnified by a massive foreground galaxy cluster lens at $z = 0.44$, modeling with Lenstool indicates a lensing amplification factor of $11\pm 2$. We determine an intrinsic rest-frame 8-1000-$\mu$m luminosity, $L_{\rm IR}$, of  $(1.3 \pm 0.1) \times 10^{13}$ $L_\sun$, and a likely star-formation rate (SFR) of $\sim 500-2000$ $M_\sun yr^{-1}$. The SED shows a remarkable similarity with the Cosmic Eyelash from optical-mid/IR to sub-millimeter/radio, albeit at higher fluxes.

\end{abstract}

%% Keywords should appear after the \end{abstract} command. 
%% See the online documentation for the full list of available subject
%% keywords and the rules for their use.

%% From the front matter, we move on to the body of the paper.
%% Sections are demarcated by \section and \subsection, respectively.
%% Observe the use of the LaTeX \label
%% command after the \subsection to give a symbolic KEY to the
%% subsection for cross-referencing in a \ref command.
%% You can use LaTeX's \ref and \label commands to keep track of
%% cross-references to sections, equations, tables, and figures.
%% That way, if you change the order of any elements, LaTeX will
%% automatically renumber them.

%% We recommend that authors also use the natbib \cites
%% and \citet commands to identify citations.  The citations are
%% tied to the reference list via symbolic KEYs. The KEY corresponds
%% to the KEY in the \bibitem in the reference list below. 

\section{Introduction} \label{sec:intro}

Submillimeter galaxies (SMGs) provide important clues on the formation and evolution of massive galaxies in the distant universe \citep{Ca14}. These high-z dusty starbursts, peaking at redshift z=2.3-2.5 \citep{Ch05,Si14,St16} are massive and most probably the progenitors of present-day ellipticals (e.g. \cite{Iv13}). Given the typical sizes of starbursts (less than a few kpc, e.g. \cite{Ho16}) even with ALMA and {\it HST} it is difficult to achieve the desirable spatial resolution to carry out detailed studies of star-forming regions in these dusty galaxies. 

Strong gravitational lensing by massive galaxy clusters can enhance the apparent brightness of SMGs \citep{Sm97,Sm02}. Wide {\it Herschel} surveys as H-ATLAS, HerMES or the South Pole Telescope and the {\it Planck} space mission provided an increasing number of bright, lensed SMGs \citep{Ne10,Vi13,We13,Ca15,Ha16,St16} that may allow their study at a resolution of 100 pc, close to the size of GMCs, and thus understand better their ISM properties (see e.g. \cite{Vl15,Sw15}) and compare then with galaxies in the local universe. Among SMGs it is outstanding the serendipitously discovered lensed galaxy SMM J2135-0102 at z=2.3259 (the Cosmic Eyelash \cite{Sw10,Iv10,Da11}).  

In this Letter we present the first result of a search for the brightest examples in the sky of dusty star-forming high redshift galaxies using data of the {\it Planck} and {\it WISE} space missions. We report the discovery of {\it WISE} J132934.18+224327.3 as a  magnified SMG at $z=2.044$ with spectral energy distribution (SED) remarkably similar tothat of the Cosmic Eyelash but higher apparent brightness at all wavelengths from visible to radio.
We adopt a flat $\Lambda$CDM cosmology from \cite{Pl14} with $H_0=68$ km s$^{-1}$ Mpc$^{-1}$, $\Omega_m=0.31$ and $\Omega_\Lambda = 1 - \Omega_m$ 

\section{The search} \label{sec:search}

In order to find bright lensed SMGs in the full-sky, we carried out  a cross-matching between the 
{\it AllWISE}\footnote{\url{http://wise2.ipac.caltech.edu/docs/release/allwise/}} and 
{\it Planck}\footnote{\url{http://pla.esac.esa.int/pla/\#home}} full-sky compact source catalogs. As a reference we adopted the SED of the Cosmic Eyelash 
\citep{Sw10}, from the MIR to the submillimeter. Our aim was to identify candidates in these catalogs with MIR colors similar to the Cosmic Eyelash and 
strong submillimeter fluxes. First we built a full-sky selection of galaxies verifying the color criteria $W1-W2>0.8$, $W2-W3<2.4$, 
$W3-W4>3.5$ and detection at $S/N>3$ in the four bands of the {\it AllWISE} catalog, we limited the sample to galactic latitude b $\geq$20$^{o}$. 
These color criteria were employed by \citep{Ig17} in their NIR/MIR search for lensed submillimeter galaxies. 
These criteria also select SMGs with SED similar to the average SED of 73 spectroscopically identified SMGs from \cite{Mi10,Ha11} and to the ALESS SMG 
composite SED \citep{Sw14}. Only SMGs at $z=2-2.8$ are expected to fulfill these color conditions.

Then, we requested the selected objects to have a {\it Planck} source detected within 5 arcmin with submillimeter flux ratios consistent with those expected for SMGs at $z=2-3$. We found 8 {\it WISE} candidates with SEDs consistent with the adopted references, however the {\it Planck} sub-millimeter/millimeter maps for most of these sources clearly show contamination by galactic dust emission. Inspection of the {\it Planck} maps and available images from different optical and near-IR catalogs identified {\it WISE} J132934.18+224327.3, as the most likely counterpart of the submillimeter {\it Planck} source PCCS2 857 G007.94+80.29 and very promising candidate to an ultra-bright SMG analog to the Cosmic Eyelash.  

A subsequent search in CADC\footnote{\url{http://www.cadc-ccda.hia-iha.nrc-cnrc.gc.ca/en/}} 
provided detections by SCUBA-2 at James Clerk Maxwell Telescope of a sub-millimeter source with coordinates consistent within 1 arcsec with those of {\it WISE} J132934.18+224327.3  which could be responsible of the observed {\it Planck} submillimeter fluxes. 
In fact, the position of the source coincides with a strong lensing cluster SDSS 1329+2243 at $z=0.44$ \citep{Bo14} which has been observed by \cite{Jo15} with JCMT/SCUBA-2 and reported a source in snapshot observations at 850 and 450 $\mu$m  with fluxes S$_{450}\approx$605 mJy and S$_{850}\approx$130 mJy, 
consistent with being the main counterpart of the {\it Planck} source. \cite{Jo15} reported arc structures in Keck images suggesting it could be a 
lensed submillimeter source. Additional detections in the radio band are found in FIRST\footnote{\url{http://sundog.stsci.edu/first/catalogs.html}}.

We used Vizier at CDS\footnote{\url{http://vizier.u-strasbg.fr/viz-bin/VizieR}} to find the most likely counterpart of the {\it WISE/Planck/SCUBA-2} source in the near-IR and visible. In {\it HST-ACS}\footnote{\url{http://archive.stsci.edu/}} images a lensed galaxy is also detected at less than 1 arcsec of {\it WISE} J132934.18+224327.3. 
We postulate this lensed galaxy is the optical counterpart of the strong submillimeter and mid-IR source (see Fig.~1 and Fig.~2a). 
Hereafter we designate this lensed galaxy  as `Cosmic Eyebrow`.

\begin{figure}
\plotone{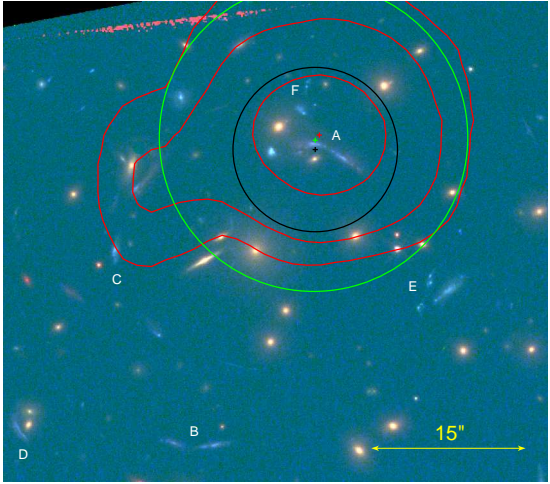}
\caption{
RGB-image of Cosmic Eyebrow with the {\it HST/WFC3} filters, blue F390W, green F606W and red F160W. North is up, East is left.
Red contours are {\it WISE} channel 4 and black (green) circles represent SCUBA-2 450 (850) $\mu$m detections. Crosses are the centroids of the
{\it WISE} and SCUBA-2 detections and they are within 1 arcsec from the {\it HST} main source of the arc. Capital letters indicate 
the six families of multiply lensed background galaxies used to perform a mass reconstruction of the foreground galaxy cluster (see text). 
}
\end{figure}

\section{Observations and data reduction} \label{sec:obser}

\subsection{GTC Spectroscopy} \label{sec:gtc}

We have obtained spectroscopy of the lensed galaxy with the optical imager and spectrograph OSIRIS at the 10.4 m Gran Telescopio de Canarias (GTC), on the night of 2017 April 18 in clear conditions and dark moon with $0.8^{\prime\prime}$ seeing. Observations were made using the low-resolution, long-slit mode with a $0.8^{\prime\prime}$ 
slit and the R500B grism. In this configuration the resolution is $R\sim 540$ and we have 3.54 $\AA$/pix from $\sim $ 3700 to 7200 $\AA$.
The slit was oriented along the line of the arc structure of the lensed source (P.A. of 64.22$^{\rm o}$). We grouped observations in 
two observing blocks of two exposures of 1385 seconds each with a 60 second acquisition image per block in $g$-band (see Fig.~2). The seeing limited image shows an arc-like galaxy in the expected position for which we measured  $g_{\rm AB}=22.85 \pm 0.03$ for region 1 and $g_{\rm AB}=23.43 \pm 0.04$ for region 2. 
The spectroscopic observations, with a total integration time of 5540 s, were reduced using the noao/twodspec and noao/onedspec packages of IRAF to yield fluxed, 
wavelength-calibrated spectra. We extracted spectra for the full arc region along the slit  and separately for  encircled regions 1 and 2. 
Our GTC spectroscopy reveals that the two regions in the arc are the same source. In Fig.~3 we show the individual spectra for each arc region 
and the total arc spectrum. The spectra of the two arc regions show  strong absorption features. We derive the redshift from the fit of 
well measured lines of SiII $\lambda 1260.4$, OI $\lambda 1302$, CII $\lambda 1334.5$, SiII $\lambda 1526.7$ and AlII $\lambda 1670.7$, 
for the full arc we find  $z= 2.0448 \pm 0.0004$.

\begin{figure}
\gridline{\fig{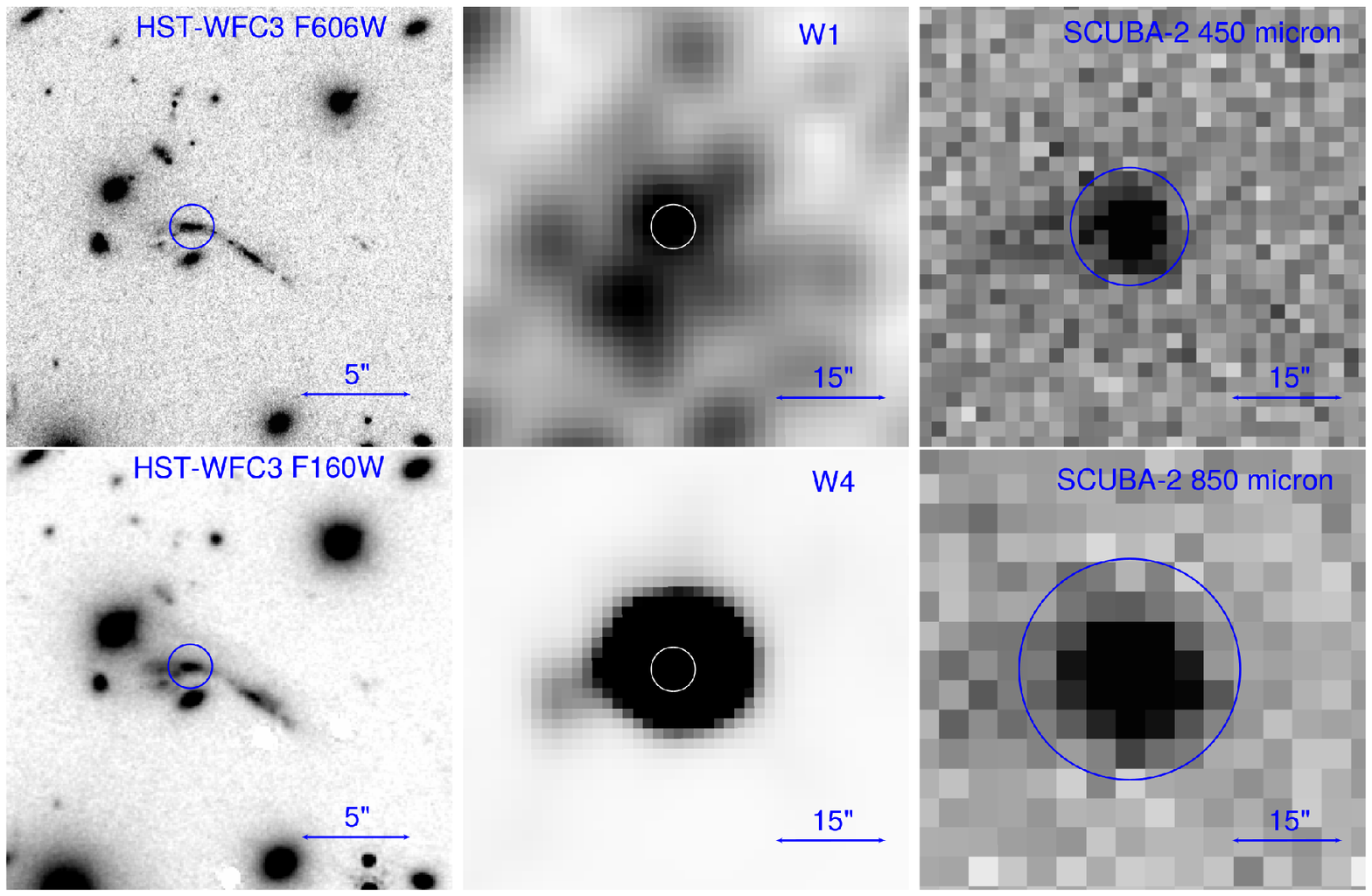}{0.27\textwidth}{(a)}
          \fig{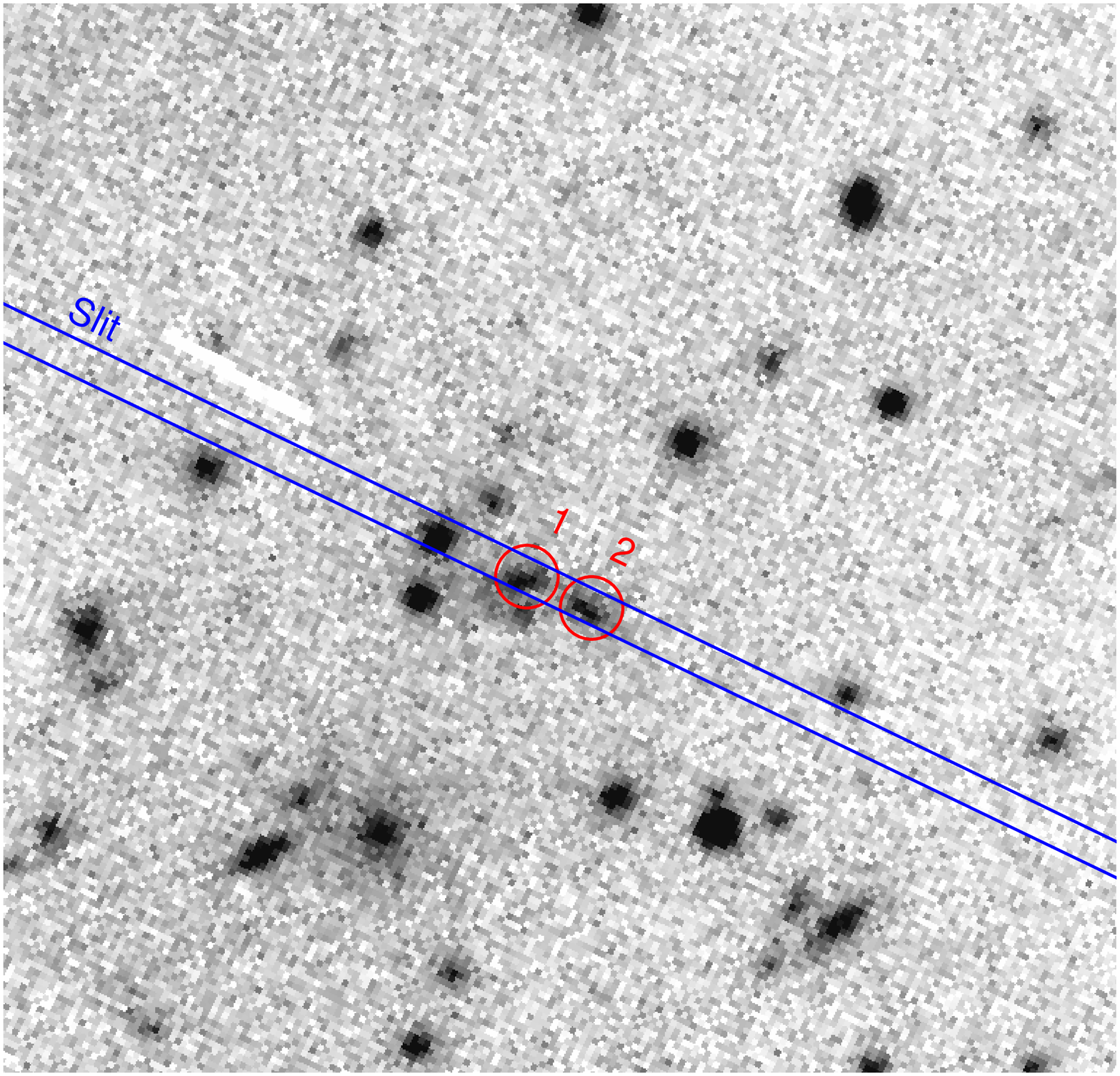}{0.18\textwidth}{(b)}
          }
\caption{ 
(a) Imaging of Cosmic Eyebrow with {\it HST} (left column), {\it WISE} (middle column) and SCUBA-2 (right column). The circles represent the counterpart of the 
{\it Planck} source in the related imaging. North is up, East is left.
(b) GTC $g$-band imaging of Cosmic Eyebrow. We show the orientation of the long-slit (slit width of 0.8$^{\prime\prime}$)  for our spectroscopic observations. Spectra were extracted from the two encircled regions of the arc. North is up, East is left.
}
\end{figure}

\subsection{Gemini Spectroscopy} \label{sec:gemini}

Archival GEMINI NIR-spectroscopic observations were found in CADC for region 1 of the arc. The spectrum was obtained using the cross-dispersed (XD) mode 
of the Gemini Near-IR Spectrograph on the 8.1 m Gemini North telescope on the night of 2014 May 10 (under PI: Jane Rigby program ID: XGN-2014A-C-3) 
with the "short blue" camera, 32 l/mm grating 
and $0.68^{\prime\prime}$ slit. The slit used in this XD mode is  $7^{\prime\prime}$ in length and
the orientation was with P.A. of 89.7$^{\rm o}$ in region 1. The telescope was nodded (typically $3.5^{\prime\prime}$ distant) in an ABBA-type pattern.
We took the raw and calibration data from the CADC and reduced them using the Gemini IRAF package 
version v1.13.1 and following \cite{Ms15} and the instruction on the Reducing XD spectra webpage from 
GEMINI observatory\footnote{\url{http://www.gemini.edu/sciops/instruments/gnirs}}.
The total integration time for the spectrum was 3600 seconds and two telluric stars of spectral type A0V were observed immediately before and after of the galaxy. 
Only in the K band we have found useful data with  a clear detection of H$_{\alpha}$ in emission. 
In the inset of Fig.~3, we show the spectrum near the $H_\alpha$ emission line from which we obtain a redshift of $z=2.0439 \pm 0.0006$ 
in agreement with the rest-frame UV-spectrum. This redshift is consistent with the value  published in \cite{Og12} ($z\approx2.04$) 
for a source coincident in position with the lensed galaxy.

\begin{figure}
\plotone{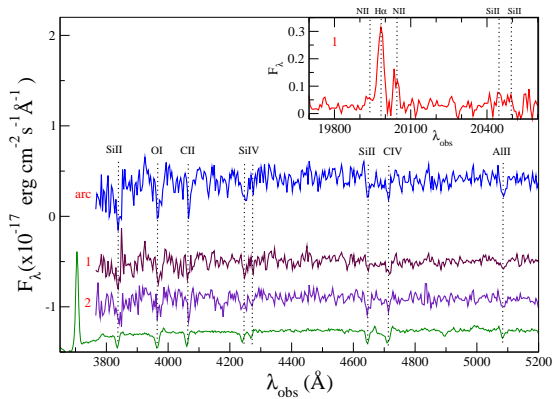}
\caption{
GTC/OSIRIS spectra of the two regions (up) and for the each individual regions (both down shifted for clarity). 
We mark the positions of the absorption features identified in the spectra. For comparison, we show the composite spectrum of LBGs from \cite{Sh03}.
We also show in the inset the Gemini spectrum for region 1 and mark the emission lines we identify, 
the $H_\alpha$ emission line yields a redshift for the source of $z=2.0439$.
}
\end{figure}

\section{Analysis and discussion} \label{sec:dis}

\subsection{Lens modeling} \label{sec:lens}

We have used Lenstool\footnote{\url{https://projets.lam.fr/projects/lenstool/wiki}} \citep{Kn93, Ju07} to perform a mass reconstruction of the foreground cluster, assuming a parametric model for the distribution of dark matter. This model was constrained using the location of the multiple images identified in the cluster. We used a simple model with a single cluster-scale mass component, as well as individual galaxy-scale mass components centered on each cluster member. For each component, we used a dual pseudo-isothermal elliptical mass distribution (dPIE, also known as a truncated PIEMD, \cite{Li05}). Sextractor and visual inspection 
of the {\it HST/WFPC3} images provided an identification of six families of multiply lensed background galaxies, arclets with 2-6 components each, 
based on proximity, colors and shape. The redshift of the family of Cosmic Eyebrow (A in Fig.~1) was fixed at $z=2.044$, 
and for the other families was set as free parameter. 
When modeling the lensing, the redshift of 3 families (B, C and E in Fig.~1) was found very close to $z=2.044$, so we fixed it to $z=2.044$ 
and only for 2 families the redshifts were considered  a free parameter. We checked the model with the reconstruction of the arclets, and found good agreement between the 
positions for image arcs and the reconstruction of each component of the families. The data are best fitted with a cluster-scale potential 
of ellipticity $e = 0.226$, position angle PA =47.00$^{\rm o}$, and velocity dispersion $\sigma_{\rm PIEMD} =830$ km/s.
The enclosed mass within an aperture of 250 kpc is $M=1.8 \pm 0.5  \times 10^{14}M_\sun$ with an Einstein radius of 
$\theta_{\rm e}= 11. \pm 0.4 ^{\prime\prime}$ at $z=2.044$. The amplification is obtained from the comparison between 
the main characteristic of the galaxies in the source plane and in the image plane, we find an amplification factor of $11\pm 2$ 
for the two main members of the Cosmic Eyebrow family, which are our spectroscopy regions on the arc indicated in Fig.~2. 
The mean amplification factors for each family with $z=2.044$ are A $\sim 11\pm 2$, B $\sim 7\pm 2$, C $\sim 14\pm 2$, E $\sim 7\pm 2$, the largest amplification is for
the brightest member of family C $\sim 20\pm 2$, for the other two families the mean amplification factors are D $\sim 4\pm 2$ ($z\sim 2.9$) and 
F $\sim 5\pm 2$ ($z\sim 1.0$).

\subsection{SED} \label{sec:sed}

We plot the SED (Fig.~4) assuming the lensed source in the {\it HST} and GTC images is the counterpart of the SCUBA-2 and {\it WISE} detections 
(all  positions coincident within 1 arcsec) and list the photometry data in table 1. We calculate the upper limit for $K_s$ and $H$ bands from UKIDSS 
catalog\footnote{\url{http://wsa.roe.ac.uk/}} and for
60 and 100 $\mu$m from {\it IRAS} Sky Survey Atlas\footnote{\url{http://irsa.ipac.caltech.edu/applications/IRAS/ISSA/}}.
This SED is consistent with a source at redshift $z=2-2.5$ and, as expected, it is very well fitted by the SED of the Cosmic Eyelash in that redshift range. At all frequencies from optical to radio, Cosmic Eyebrow is brighter than the Cosmic Eyelash. 

We calculate the rest-frame 8-1000-$\mu$m luminosity, $L_{\rm IR}$, from direct integration of the data fit, and obtain
an intrinsic luminosity for our galaxy $L({\rm Eyebrow})= (1.3 \pm 0.1) \times 10^{13}$ $L_\sun$,
indicating a SFR of $\sim 2000$ $M_\sun yr^{-1}$ \citep{Ke98}, which assumes a Salpeter IMF, assuming a Chabrier IMF would possibly be a 
factor of $\sim$ 1.8 lower \citep{Ca14}.

\begin{figure}
\plotone{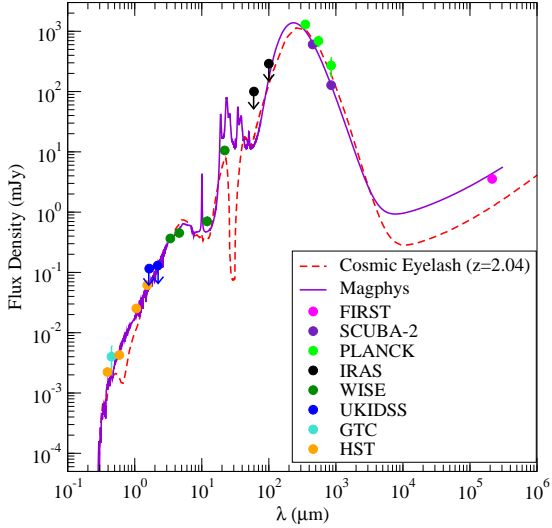}
\caption{
Multi-wavelength spectral energy distribution of Cosmic Eyebrow. The SED of Cosmic Eyelash shifted to $z=2.044$ is consistent with the measurements of the 
galaxy at all frequencies from the optical to the radio. Due to low-spatial resolution, most probably the FIRST measurement in the radio regime has 
been contaminated by nearby source(s). 
}
\end{figure}

We also use the MAGPHYS code that allows us to fit simultaneously the ultraviolet-to-radio SED, so we can constrain physical parameters \citep{Cu15}. 
The fit is shown in (Fig.~4), we would benefit from measurements in   photometric bands between 22 and 350 $\mu$m in order to
establish which SED describes better the Cosmic Eyebrow. From this fit we give the median-likelihood estimates (and confidence ranges) 
of several physical parameters corrected for lensing amplification, stellar mass $log(M_\ast/M_\sun) = 11.49^{+0.06}_{-0.05}$, 
star formation rate $log(SFR/M_\sun yr^{-1}) = 2.73^{+0.02}_{-0.01}$, mass-weighted age
$log(age_M/yr) = 9.28^{+0.01}_{-0.01}$, average V-band dust attenuation $A_V = 3.30^{+0.02}_{-0.02}$, H-band mass-to-light ratio 
$log(M_\ast/L_H) = 0.14^{+0.01}_{-0.01}$, total dust luminosity $log(L_{\rm dust}/L_\sun) = 13.27^{+0.01}_{-0.02}$, 
luminosity-averaged dust temperature $T_{\rm dust}/K = 40^{+7}_{-2}$ and total dust mass $log(M_{\rm dust}/M_\sun) = 9.19^{+0.15}_{-0.08}$. 
This physical parameters compare well with the average properties of the ALESS SMGs in \cite{Cu15}. 
The SFR obtained with MAGPHYS is lower than estimated with the \cite{Ke98} relation. 
This is partially due to the use of the Chabrier IMF and to the additional dust heating caused by the relatively old stellar 
populations of our galaxy which increases the dust luminosity at fixed SFRs \citep{Cu15}.

\begin{deluxetable}{lcc}[b!]
\tablecaption{Photometry \label{tab:photometry}}
\tablecolumns{3}
\tablenum{1}
\tablewidth{0pt}
\tablehead{
\colhead{Wavelength($\mu$m) } &
\colhead{Flux (mJy)\tablenotemark{a,b}} &
\colhead{Observatory/Instrument}  \\
}
\startdata
0.3921 &  0.0022 $\pm 0.0001$ & {\it HST/WFC3} \\
0.45 &  0.004 $\pm 0.002$ &  GTC/OSIRIS \\
0.5887 & 0.0043 $\pm 0.0001$ & {\it HST/WFC3} \\
1.0552 & 0.0252 $\pm 0.0004$ & {\it HST/WFC3}  \\
1.5369 & 0.0614 $\pm 0.0006$ & {\it HST/WFC3} \\
1.644 & $<$ 0.112 \tablenotemark{c}  &  UKIDSS \\
2.199 & $<$ 0.110  \tablenotemark{c} &  UKIDSS \\
3.4 & 0.37 $\pm 0.01$ & {\it WISE} \\
4.6 & 0.45  $\pm 0.02$ & {\it WISE} \\
12 & 0.7 $\pm 0.1$ & {\it WISE} \\
22 & 10.6 $\pm 0.8$ & {\it WISE} \\
60 & $<$ 100 \tablenotemark{c} & {\it IRAS} \\
100 & $<$ 300 \tablenotemark{c} & {\it IRAS} \\
450 & 604 $\pm 86$ & SCUBA-2 \\ 
850 & 127 $\pm 11$ & SCUBA-2 \\
350 & 1298 $\pm 200$ & {\it Planck} \\
550 & 692 $\pm 100$ & {\it Planck} \\
850 & 271 $\pm 90$ & {\it Planck} \\
21.4\tablenotemark{d} & $ 3.56 \pm 0.14$ & FIRST \\ 
\enddata
\tablenotetext{a}{Uncorrected for lensing amplification.}
\tablenotetext{b}{Total arc flux.}
\tablenotetext{c}{Flux upper limit.}
\tablenotetext{d}{Wavelength in cm.}
%\tablecomments{
%}
\end{deluxetable}

\subsection{Spectroscopy} \label{sec:spectdis}

The  UV rest-frame  spectrum of the Cosmic Eyebrow resembles that of 
Lyman break galaxies (LBG) \citep{Sh03} and display the strong interstellar absorption features typical of the spectra of
starburst galaxies \citep{Ca13}. We identify in the observed spectrum low-ionization resonance interstellar metal lines such as
SiII $\lambda 1260.4$, OI 1302+Si II $\lambda1304$, CII  $\lambda 1334$, SiII $\lambda 1526$ and AlII $\lambda 1670$ which are associated with the
neutral interstellar medium. We also detect  high-ionization metal lines of Si IV $\lambda\lambda 1393,1402$ and CIV $\lambda\lambda 1548, 1550$ which are associated with
ionized interstellar gas and PCygni stellar wind features.  The CIV feature exhibits a strong interstellar absorption component plus a 
weaker  blueshifted broad absorption and marginal evidence of redshifted emission associated with stellar winds which likely originate in
main-sequence, giant and supergiant O stars \citep{Wa84}. Our data do not show evidence for the  He II $\lambda 1640$ which could be an indication of a low ratio of Wolf-Rayet to O stars \citep{Scha98}.

In the spectrum of the region 1 of the arc we find $H_\alpha$ at $19982 \pm 4$ $\AA$ and the [NII] $\lambda 6583$ at $ 20041 \pm 5$ $\AA$. 
Full width at half maximum (FWHM) line widths are $386 \pm 20$ km s$^{-1}$ and $331 \pm 20$ km s$^{-1}$, after correction for instrumental 
profile we estimate the intrinsic width of these lines is $150 \pm 25$ km s$^{-1}$. 
The low [NII]$\lambda 6583$/$H_\alpha$ ratio of $0.33 \pm 0.04$ suggests a star-forming region \citep{Ke13}. 
The EW$_{rest}$ ($H_\alpha$)= $111 \pm 10$ $\AA$  compares well with values reported in other well known SMGs of similar luminosity \citep{Ol16}. 

\section{Conclusions} \label{sec:concl}

A cross-match of the  {\it AllWISE} and {\it Planck} compact source  catalogs aimed to identify the brightest examples of submillimeter galaxies in the full sky uncovered {\it WISE} J132934.18+224327.3, the Cosmic Eyebrow, a $z=2.0439 \pm 0.0006$ galaxy with spectral energy distribution from mid-IR to submillimeter similar to that of the Cosmic Eyelash and  higher observed fluxes. Archival data of SCUBA-2 and {\it HST} reveals a multiple lensed galaxy at the  position of the strong submillimeter source. Follow-up observations with GTC/OSIRIS provided a precise redshift determination  of the lensed galaxy and a rest-frame UV spectrum with clearly identified low-ionization resonance interstellar metal lines which is consistent with that of starburst galaxies. Near-IR spectroscopy with Gemini-North showed H$_{\alpha}$ in emission at a strength consistent with values reported for other SMGs. This galaxy is gravitationally magnified by a massive cluster at $z = 0.44$, modeling with Lenstool indicates a lensing amplification factor of $11\pm 2$. The intrinsic rest-frame 8-1000-$\mu$m luminosity of the lensed galaxy is $(1.3 \pm 0.1) \times 10^{13}$ $L_\sun$ indicating a likely SFR of $\sim 500-2000$ $M_\sun yr^{-1}$. 
The SED of this new SMG resembles the Cosmic Eyelash from optical-mid/IR to sub-millimeter/radio, albeit at higher intrinsic luminosity. 
It is one of the brightest examples of SMGs so far reported which may enable detailed high spatial resolution studies of star-forming regions in such dusty galaxies.

\acknowledgments

Based on observations made with the GTC telescope, in the Spanish Observatorio del Roque de los Muchachos of the Instituto de 
Astrof\'isica de Canarias. This work has been partially funded by projects ESP2015-69020-C2-1-R, ESP2014-56869-C2-2-P,
and AYA2015-69350-C3-3-P (MINECO). H.D. acknowledges financial support from the Spanish Ministry of Economy and 
Competitiveness (MINECO) under the 2014 Ram\'on y Cajal program MINECO RYC-2014-15686.
We thank David Valls-Gabaud for his suggestions and comments on various aspects of this work. 
We are grateful to the anonymous referee for his useful comments on the paper.

%% To help institutions obtain information on the effectiveness of their 
%% telescopes the AAS Journals has created a group of keywords for telescope 
%% facilities.
%
%% Following the acknowledgments section, use the following syntax and the
%% \facility{} or \facilities{} macros to list the keywords of facilities used 
%% in the research for the paper.  Each keyword is check against the master 
%% list during copy editing.  Individual instruments can be provided in 
%% parentheses, after the keyword, but they are not verified.

%\vspace{5mm}
%\facilities{GTC(OSIRIS), GEMINI, HST(WFC3), SCUBA-2}

%% Similar to \facility{}, there is the optional \software command to allow 
%% authors a place to specify which programs were used during the creation of 
%% the manusscript. Authors should list each code and include either a
%% citation or url to the code inside ()s when available.

%\software{IRAF, SExtractor, Lenstool }

%% This command is needed to show the entire author+affilation list when
%% the collaboration and author truncation commands are used.  It has to
%% go at the end of the manuscript.
%\allauthors

%% Include this line if you are using the \added, \replaced, \deleted
%% commands to see a summary list of all changes at the end of the article.
%\listofchanges

\end{document}